28th International Conference on Knowledge-Based and Intelligent Information & Engineering Systems (KES 2024)

# Sustainable Digitalization of Business with Multi-Agent RAG and LLM

Muhammad Arslan[a]*, Saba Munawar[b] and Christophe Cruz[a]

[a]Laboratoire Interdisciplinaire Carnot de Bourgogne (ICB), Université de Bourgogne, Dijon, France
[b]National University of Computer and Emerging Sciences (NUCES), Islamabad, Pakistan

**Abstract**

Businesses heavily rely on data sourced from various channels like news articles, financial reports, and consumer reviews to drive their operations, enabling informed decision-making and identifying opportunities. However, traditional manual methods for data extraction are often time-consuming and resource-intensive, prompting the adoption of digital transformation initiatives to enhance efficiency. Yet, concerns persist regarding the sustainability of such initiatives and their alignment with the United Nations (UN)'s Sustainable Development Goals (SDGs). This research aims to explore the integration of Large Language Models (LLMs) with Retrieval-Augmented Generation (RAG) as a sustainable solution for Information Extraction (IE) and processing. The research methodology involves reviewing existing solutions for business decision-making, noting that many systems require training new machine learning models, which are resource-intensive and have significant environmental impacts. Instead, we propose a sustainable business solution using pre-existing LLMs that can work with diverse datasets. We link domain-specific datasets to tailor LLMs to company needs and employ a Multi-Agent architecture to divide tasks such as information retrieval, enrichment, and classification among specialized agents. This approach optimizes the extraction process and improves overall efficiency. Through the utilization of these technologies, businesses can optimize resource utilization, improve decision-making processes, and contribute to sustainable development goals, thereby fostering environmental responsibility within the corporate sector.

---

* Corresponding author. Tel.: +33 03 80 39 50 00.
 *E-mail address:* muhammad.arslan@u-bourgogne.fr

## 1. Introduction

Businesses depend extensively on data to fuel their operations. This reliance underscores the need for efficient data utilization, which requires access to pertinent information sourced from a variety of structured, semi-structured, and unstructured channels, all customized to suit their particular industry domain. This data forms the foundation for understanding market dynamics and identifying business opportunities, enabling informed decision-making. Moreover, user engagement [1] and user-driven innovation play crucial roles in enhancing product development and market strategies. Insights gathered from diverse social media data about user and buyer experiences significantly impact product development [2], emphasizing the importance of integrating consumer perspectives into business strategies. Traditionally, businesses have employed manual methods [3], often involving business intelligence experts, to extract pertinent information from diverse data sources. However, this approach is characterized by its time-consuming nature, resource-intensive requirements, and susceptibility to human errors. In response, many companies have implemented various solutions for IE, deploying systems to drive digital transformation [4] within their organizations. Digital transformation, in essence, involves the adoption of technology to optimize business processes and enhance efficiency [4].

Over time, as the volume and diversity of data sources have expanded, companies have faced growing challenges in extracting and processing this information, requiring increased allocation of resources. This necessitates not only the upgrading of information processing systems with more powerful central processing units and memory units but also broader organizational-level initiatives for digital transformation [5]. However, while digital transformation initiatives aim to enhance business information systems and market understanding, questions arise regarding their sustainability and alignment with the UN's SDGs [6]. Sustainability refers to the ability of these initiatives to meet present needs without compromising the ability of future generations to meet their own needs [7]. The deployment of machines and the subsequent increase in carbon footprint raise significant environmental concerns, casting doubts on the long-term sustainability of current digital transformation practices.

The advent of advanced technologies like LLMs represents a significant step towards sustainable digitalization in business operations [8]. These models, equipped with the ability to comprehend and generate human-like text, offer an accessible solution for IE and processing tasks [9]. They come pre-trained on extensive datasets, providing companies with an off-the-shelf solution for tasks such as extracting relevant business data from large datasets based on contextual understanding [9]. LLMs have advantages but are often trained on general datasets and might not be tailored for specific company requirements initially [10]. To tailor these models to specific contexts, additional contextual information must be incorporated. This process ensures that LLMs can effectively address the unique requirements of individual businesses, maximizing their utility in sustainable digital transformation initiatives.

With the integration of RAG [10], enhancing LLMs through external data retrieval, businesses can automate IE from diverse sources. Multi-Agent RAG, an architecture for RAG, further refines this process by dividing IE tasks among specialized agents, each programmed to perform specific tasks and communicate with others [11, 12]. Unlike single-model approaches, Multi-Agent RAG divides responsibilities across different agents, leveraging diverse datasets. This asynchronous operation improves relevance, latency, and coherence in conversational systems. This approach enables businesses to access a wider range of insights, mitigating manual method limitations. Leveraging Multi-Agent RAG and LLMs, businesses optimize resource utilization, improve decision-making, and contribute to sustainable development goals.

The paper is structured as follows: Section 2 provides background information, Section 3 details the Multi-Agent RAG with LLM use case, Section 4 offers discussion points, and Section 5 concludes the paper.

## 2. Background

Businesses rely on data for informed decisions [1]. IE methods play a crucial role by extracting actionable insights from various data sources and enhancing decision-making [13]. The emergence of advanced language models like Bidirectional Encoder Representations from Transformers (BERT) [14] and Generative Pre-trained Transformer (GPT) [15] has further revolutionized this process by automating IE tasks, thereby enabling businesses to extract valuable insights at scale. Recent studies highlight the critical role of IE from business documents in various domains, emphasizing its importance in enhancing operational efficiency. Employing BERT-based models,

Douzon et al. [16] showcased improvements in IE accuracy, while Geletka et al. [17] extended IE capabilities through multi-modal transformers. Hamdi et al. [18] focused on extracting structured data from invoices using BERT-based approaches, addressing the challenge of unstructured data formats. Nguyen et al. [19] highlighted the robustness of BERT-based models in IE from domain-specific documents with limited data. Hedberg & Furberg [20] automated insurance policy IE, demonstrating BERT's applicability in processing complex documents. Additionally, Moreno Acevedo [21] and Zhang et al. [22] explored IE automation in business documents, further affirming BERT's effectiveness. Jacobs & Hoste [23] enabled supervised IE of company-specific events, while Dor et al. [24] focused on financial event extraction using weak supervision techniques. Pugachev et al. [25] predicted news popularity, and Hillebrand et al. [26] introduced KPI-BERT for financial reports, highlighting BERT's versatility. Bellan et al. [27] demonstrated GPT-3's efficacy in capturing business process structures, while Arslan & Cruz [28] showcased BERT's effectiveness in extracting actionable insights from business documents, collectively emphasizing the transformative potential of LLMs in IE tasks.

After analyzing the above-mentioned studies, it is clear that LLMs offer numerous benefits to businesses by extracting relevant information for business processes. In many of the systems discussed, LLMs are trained using company-specific datasets, which demand extensive data, time, and resources, making their implementation costly, time-consuming, and unsustainable. Here, the concept of using pre-trained LLMs for business IE becomes valuable. With small, company-specific datasets, RAG enables LLMs to be effectively utilized for business use cases. RAG is a technique aimed at improving the accuracy and reliability of LLMs by incorporating information obtained from external sources. For further details on RAG, readers can consult Lewis' study [10]. When designing RAG with LLM solutions, various options for existing RAG models and LLMs are available in the literature.

BabyAGI [29] employs multiple LLM-based agents for task creation, prioritization, and completion. CAMEL [30] utilizes role-playing to enable chat agents to collaborate on task completion while recording conversations for analysis. Multi-Agent Debate is explored [31, 32] as an effective method to stimulate divergent thinking and enhance factuality and reasoning in LLMs. MetaGPT [33] introduces a specialized LLM application for automatic software development. AutoGen2 [11], an open-source framework, enables developers to create LLM applications with customizable conversable agents capable of operating in various modes. Moreover, CrewAI (crewai.com) offers a cohesive platform for Artificial Intelligence (AI) agents to assume roles, share goals, and collaborate effectively in diverse tasks, demonstrating the potential for sophisticated Multi-Agent interactions in AI systems.

After studying several RAG with LLM integrated solutions, we selected CrewAI for its up-to-date features, including a Multi-Agent RAG architecture with LLM support. This platform facilitates the integration of datasets of varying structures, such as structured, semi-structured, and unstructured, utilizing the LangChain framework (Langchain.com). Furthermore, CrewAI provides compatibility with various LLMs, such as Ollama [34] and OpenAI GPT [35] solutions. For the proof-of-concept presentation, we utilized the 'gpt-3.5' LLM. In the next section, we introduce our novel Multi-Agent RAG solution with LLM for extracting business insights, specifically targeting business events, a novel application not yet explored in existing literature.

## 3. Business Insights using Multi-Agent RAG and LLM

To facilitate the pre-trained LLM's reusability within sustainable business scenarios, we present a Multi-Agent RAG architecture tailored for extracting business information, particularly emphasizing business events. These events (e.g., recruitment, material acquisition, etc.) incorporate diverse elements including business individuals, company names, locations, timestamps, and contextual details (e.g., topic), enabling the generation of comprehensive business insights from the collection of new articles in our case. To enrich these business events, additional data sources such as internal company data, financial data, and consumer review data are required (see Fig. 1), available in diverse formats including Comma-Separated Values (CSV) files, Excel spreadsheets, Portable Document Format (PDF) documents, and website content. This fusion of data enables the correlation of information such as company ownership, company financial position, contact details, and market reputation with the identified business events, enhancing the depth and context of the extracted insights. To enable the event extraction from multiple diverse data sources, we used Multi-Agent RAG architecture. Leveraging the Multi-Agent functionality, various RAG-based architectures and LLMs were explored, as discussed in the previous section. Specifically, we employed CrewAI (crewai.com/) for RAG functionality in our solution design, chosen for its updated model. Additionally, OpenAI GPT (openai.com) was incorporated to showcase its integration with RAG.

To illustrate the seamless integration of Multi-Agent RAG with LLM, Algorithm 1 is presented. This algorithm serves to enable IE regarding business events (see Fig. 2), while also facilitating the enrichment of associated information from diverse datasets. Moreover, it categorizes these events based on various business topics (e.g., construction, recruitment, acquisition, production, etc.). Algorithm 1 initializes several agents, which are autonomous entities programmed to execute IE tasks, make decisions, and communicate with each other. Each agent possesses distinct attributes: 'Role', defining its function within the crew; 'Goal', specifying its individual objective; and 'Backstory', providing context for its role and goal, fostering collaboration dynamics. Moreover, agents are linked with an LLM, the language model driving their operations. In our scenario, we employ three specialized agents, namely:

- *EventsCrawler*, tasked with loading news data and extracting business events with named entities (i.e., individuals, locations, dates, etc.),
- *EventsEnrichment* is tasked with leveraging financial data, internal company data, and consumer reviews to link entities with relevant data. For extracting data from various dataset formats like CSV files, Excel spreadsheets, PDF documents, and websites, we recommend LangChain (langchain.com), a framework tailored for LLM-powered applications. Its modules, such as CSVLoader, PyPDFLoader, and WebBaseLoader, streamline the extraction and indexing of textual content efficiently.
- *EventsExplorer*, aimed at categorizing business events based on their topics.

The tasks are then sequenced: gathering events data, enriching the data, and classifying the events. Finally, the crew, comprising a collaborative group of agents, is established through a sequential process, ensuring a systematic and organized approach to accomplishing the set objectives. Ultimately, this algorithm allows our system to generate enriched and categorized business events, providing valuable insights for informed decision-making and strategic planning in business contexts.

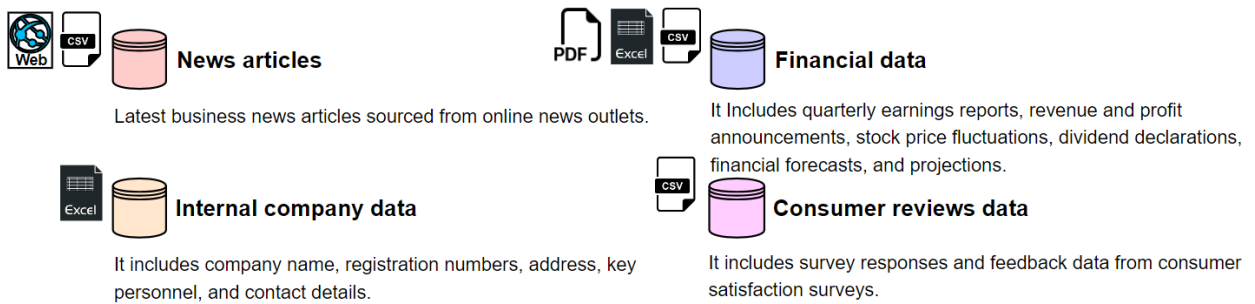

Fig. 1. Various datasets contribute to understanding business events.

## 4. Discussion

LLMs have revolutionized human-machine interactions, enabling a wide array of tasks such as IE, text generation, translation, and Question Answering (QA) [36]. However, their transformative capabilities come with a hefty computational cost, particularly during training and inference phases. To train and deploy LLMs, substantial computing power is required, typically in the form of General Processing Unit (GPU) chips [37]. Powerful LLMs necessitate thousands of GPUs for operation, each with its own environmental footprint, from manufacturing to disposal. The energy-intensive nature of GPU chip production and the vast computing resources needed for LLM training significantly contribute to their environmental impact. Studies have shown that training a single LLM can result in substantial carbon dioxide emissions, equivalent to several times the lifetime emissions of an average car or Algorithm. 1. Multi-Agent RAG with LLM for business events [12].

---

**Objective: Extracting Enriched and Categorized Business Events**

**Input:** Diverse datasets including news articles, internal company data, financial data, and consumer review data.

From **crewai** import **Crew**, **Process**, **Agent**, **Task**
From **langchain.chat_models** import **ChatOpenAI**

**Step 1:** *Initialize the LLM.*
   llm = ChatOpenAI(model='gpt-3.5')

**Step 2:** *Define the agents.*
   **EventsCrawler** = **Agent**(
    **role**='Events Crawler',
    **goal**='Load news data and extract business events with named entities.',
    **backstory**='Business news articles contain entities such as company names, individuals, contextual information, dates, and locations.',
    **llm**=llm,)
   **EventsEnrichment** = **Agent**(
    **role**='Events Enrichment',
    **goal**='Utilize Financial data, Internal Company data, and Consumer reviews data.',
    **backstory**='Associating the entities identified in the news events with their corresponding data.',
    **llm**=llm,)
   **EventsExplorer** = **Agent**(
    **role**='Events Explorer',
    **goal**='Display categorized business events.',
    **backstory**='Business events must be classified according to their topics.',
    **llm**=llm,)

**Step 3:** *Define the tasks in sequence.*
   **Crawler_task** = **Task**(description='Gather events data', agent=EventsCrawler)
   **Enrichment_task** = **Task**(description='Enrich the data', agent=EventsEnrichment)
   **Explorer_task** = **Task**(description='Classify the events', agent=EventsExplorer)

**Step 4:** *Form the crew with a sequential process.*
   **report_crew = Crew**(
    agents=[EventsCrawler, EventsEnrichment, EventsExplorer],
    tasks=[Crawler_task, Enrichment_task, Explorer_task],
    process=Process.sequential)

**Output:** Enriched and categorized business events.

---

hundreds of roundtrip flights [37, 38]. As businesses strive to meet sustainability goals, addressing the environmental impact of LLMs has become a necessary concern for sustainable technological progress.

Leveraging the reusability of LLMs emerges as a crucial strategy to reduce their environmental footprint. Rather than every business organization training its own LLM, they can capitalize on open-sourced LLMs tailored to their business operations. Integrating pre-existing LLMs with RAG technology empowers businesses to customize these models for specific requirements, enabling efficient extraction of business insights from diverse data sources with minimal resource consumption. For instance, this article demonstrates how the fusion of Multi-Agent RAG with LLMs, particularly in business event extraction, underscores the importance of LLM reusability in sustainable business practices. The proposed solution for extracting business events, leveraging Multi-Agent RAG with LLM technology, offers significant benefits in streamlining information retrieval and decision-making processes within enterprises.

Firstly, by harnessing the power of advanced LLMs, the system can efficiently analyze vast amounts of structured, semi-structured and unstructured data from diverse sources, including news articles, internal company records, financial reports, and consumer feedback. This capability enables businesses to stay updated on relevant industry developments, emerging trends, and market dynamics in real-time, facilitating proactive decision-making. Additionally, the designed system enhances operational efficiency by automating the extraction and synthesis of critical business insights, reducing manual effort and time required for data analysis. Moreover, by providing comprehensive and accurate information on business events, the system empowers stakeholders to make informed strategic decisions, identify growth opportunities, mitigate risks, and optimize resource allocation effectively.

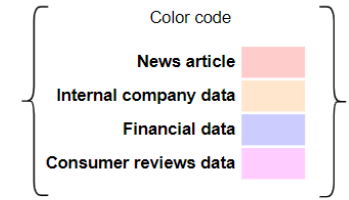

Fig. 2. Illustrating the process of extracting business events from news articles. Initially, the news articles are loaded, followed by the extraction of events, their enrichment with diverse data sources, and ultimately, the exploration of categorized events by the analyst.

The presented business system serves as a QA system, offering business experts a unified interface to leverage diverse data sources for a comprehensive understanding of the business landscape. The resulting output from this system can be utilized to develop various types of business dashboards, demonstrated in Fig. 3. This business dashboard provides an extensive overview of business events spanning various categories sourced from news articles, including Urban Planning, Renewable Energy, Photovoltaic, Fundraising, Recruitment, Acquisition, Agrivoltaics, Production, Wind power, and Healthcare, within a specified timeframe. Analysts can efficiently visualize the distribution of these events by category through a graphical representation (located in the bottom-right section of Fig. 3). Furthermore, analysts can refine their focus by filtering and extracting specific events from chosen categories, such as Photovoltaic events, tailored to their preferences (visible in the top-right portion of Fig. 3). Additionally, the dashboard interface furnishes a geographical heatmap (found in the top-left corner of Fig. 3), highlighting the geographical concentration of Photovoltaic events across different regions of the country. Lastly, users can explore the companies driving these Photovoltaic events, with pertinent information presented in the bottom-left section of Fig. 3. This multi-dimensional approach empowers analysts to explore invaluable insights about industry trends, geographical patterns, and key players shaping business events within specific sectors.

Beneath the surface of these insightful business dashboards, a sophisticated network of diverse datasets fuels their operation. In the dynamic business realm, where new technologies and concepts emerge continually, datasets experience rapid and frequent transformations. The integration of Multi-Agent RAG with LLM offers a seamless plug-and-play solution, enabling effortless swapping of outdated datasets with fresh, current ones sourced from news articles. Leveraging cutting-edge LLM models, companies can smoothly implement advanced Business Intelligence (BI) solutions, bypassing the need for extensive Research and Development (R&D) endeavors. This streamlined approach not only lightens the load on company resources but also particularly saves time, empowering organizations to maintain agility and adaptability during the dynamic shifts of the market landscape.

## 5. Conclusion

Businesses heavily rely on data for operations and decision-making, but building custom LLMs for information processing is costly and environmentally taxing. Existing pre-trained LLMs provide a cost-effective solution, transforming human-machine interactions. Integrating these reusable LLMs with RAG technology is crucial for sustainability, offering tailored insights while reducing environmental impact. By tailoring LLMs to specific business needs and employing a Multi-Agent RAG architecture, companies can optimize insight extraction processes, improve decision-making, and contribute to sustainable development goals. While this research study offers numerous benefits, it is essential to acknowledge its limitations. Firstly, the dataset of news articles utilized is limited to one month's worth of data, which may restrict the scope and depth of analysis. Furthermore, the absence of social media data, a significant source of real-time information, represents a notable gap in the study. To address these limitations, future research endeavors should focus on expanding the dataset to include a more extensive timeframe and incorporate social media data for a comprehensive analysis. However, expanding the dataset may lead to increased response times by the RAG, necessitating further investigation into optimizing system performance. Additionally, while this study utilized the LLM by OpenAI for proof-of-concept demonstration, exploring and evaluating other latest LLM models could provide valuable insights into their effectiveness within the business domain. Thus, future research should aim to compare and assess various LLM models to identify the most suitable ones for business IE applications.

**Dataset Availability**

https://drive.google.com/file/d/18UB-TamXvCFpq0edfH7EVPjBl9Ec34dC/view?usp=sharing

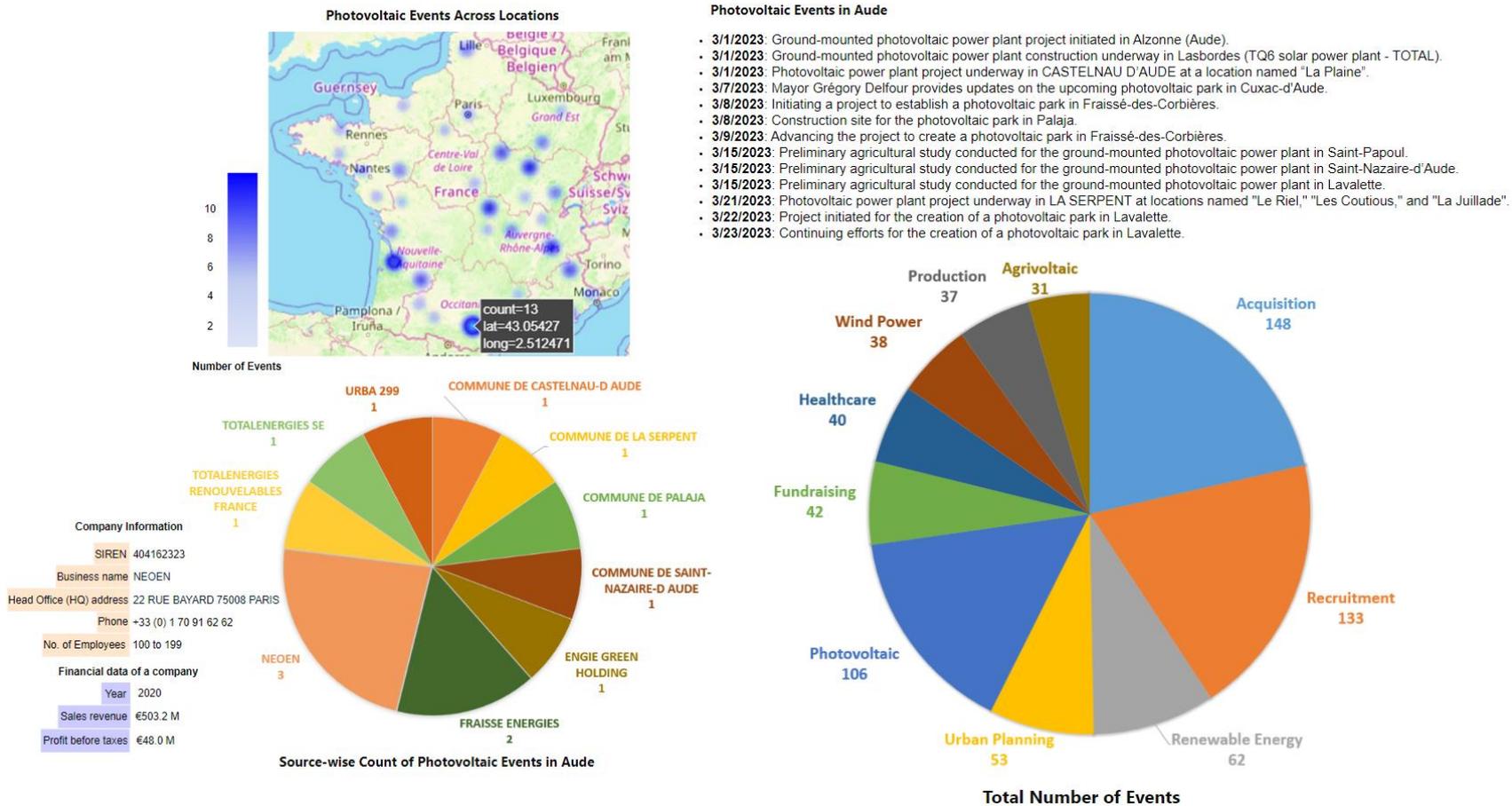

Fig. 3. A potential layout for a business dashboard constructed using insights derived from the system's multi-source data extraction capabilities: It will give the quick look of number of business events belonging to different categories happening at a particular month (below-right). Based on the analyst's preference, the related events belonging to the certain category of business (e.g. Photovoltaic events) can be extracted (see Top-right). Moreover, the location map (top-left) will show the density of "Photovoltaic events" emerging from certain geographic locations across the country. Lasty, the analysts can explore the companies (bottom-left) which are originating these Photovoltaic events business events.

**Acknowledgment**

The authors thank the French government for the funding.